\documentstyle[preprint,aps]{revtex}
\begin{document}

\draft

\title
{ Fermion Back-Reaction and the Sphaleron}
\author{Andr\'e Roberge\cite{email}}
\address{D\'epartement de physique et d'astronomie,
Universit\'e Laurentienne,
Sudbury, Ontario, Canada, P3E~2C6\\
}
\maketitle

\begin{abstract}
Using a simple model, a new sphaleron solution which
incorporates finite
fermionic density effects is obtained.  The main result is
that the height of
the potential barrier (sphaleron energy) decreases as the
fermion density
increases.  This suggests that the rate of sphaleron-induced
transitions
increases when the fermionic density increases.  However the
rate increase is
not expected to change significantly the predictions from the
standard
sphaleron-induced baryogenesis scenarios.
\end{abstract}
\pacs{11.15.Kc,11.10.Kk}
\vskip 2cm

Nonconservation of baryon number in the standard model through
quantum effects
 (anomalies) is well known\cite{tHooft}.
 While instanton mediated baryon decays are negligible, the
same cannot be said
for
 transitions occuring because of monopole
catalysis\cite{Callan-Rubakov},
 high temperature\cite{Kuzmin,Khlebnikov}, high
densities\cite{Rubakov1,Rubakov2}, or in the presence of a
heavy
particle\cite{Rubakov3} (for an excellent review of these four
mechanisms, see
ref.~\cite{Matveev88}).

The basis for all these transitions is the level crossing
phenomenon\cite{Christ} which is usually illustrated  by
looking at adiabatic
changes in the gauge field configuration and at the
accompanying  variation in
the energy levels for the fermions resulting in a change in
the fermion number.
 This description neglects the effect of the fermion
back-reaction: a change in
the fermion density can introduce a change in the gauge field
configuration,
just as a change of gauge field    configuration can change
the fermion
density.  This is most easily seen in the Schwinger model
where this
back-reaction of the fermions is responsible for oscillation
in the fermion
number \cite{Boyanovsky2}.
The fermion back-reaction is a purely quantum mechanical
effect being a direct
consequence of the anomaly equation.
Since  the focus of fermion number violation has been in the
study of solutions
to the classical
 equation of motion (e.g.\ instantons and sphalerons),  it is,
therefore, not
surprising that little attention has been  paid to this
back-reaction.
 Further, it might be very difficult to properly take into
account the fermion
back-reaction in realistic 3+1 dimensional theories since the
resolution of
even seemingly straightforward related issues, like the gauge
invariance of the
free energy at finite temperature and fermionic density
\cite{Redlich}, require
a careful treatment of non-perturbative effects to be properly
resolved\cite{Rutherford}.

Fortunately, the situation is simpler in 1+1 dimensional
models where one can,
through bosonization  \cite{Boyanovsky1},
 take into account the fermion back-reaction at the classical
level.
The present work illustrates some of  the non-trivial effects
of this fermion
back-reaction using
a simple model which has been extensively studied in the past,
namely
 the Abelian Higgs model axially coupled to fermions
 (see for example
\cite{Rubakov2,Krasnikov,Matveev87,Roberge90,Bochkarev,Carson})
{}.

The Lagrangian density describing this model is
\begin{equation}
   {\cal L} = \bar{\psi} i\gamma^\mu(\partial_\mu - ie\gamma^5
A_\mu)\psi
        - \frac{1}{4} F_{\mu\nu}F^{\mu\nu} + (D_\mu
\phi)(D^\mu\phi)^*
 -\lambda(|\phi|^2- c^2)^2
\end{equation}
where $D_\mu = \partial_\mu -ieA_\mu$ and the space coordinate
extends from
$-L$ to $L$.
 This model is to be regulated such that the gauged current,
$\bar{\psi}
\gamma^\mu\gamma^5 \psi$ is conserved while the vector current
obeys the
anomaly
equation
   \begin{equation}
   \partial_\mu\, \bar{\psi}\gamma^\mu\psi\equiv \partial_\mu
J^\mu =
    -\frac{e}{2\pi} \epsilon^{\mu\nu} F_{\mu\nu}.
\label{eq:anomaly}
   \end{equation}

 Instead of working directly with the fermionic Lagrangian, it
is preferable to
use the Bose-equivalent
form\cite{Boyanovsky2,Boyanovsky1,Roberge90}:
\begin{equation}
   {\cal L} = \frac{1}{2}\left(\partial_\mu \chi
              -  \frac{e}{\sqrt{\pi}} A_\mu\right)^2
       - \frac{1}{4} F_{\mu\nu}F^{\mu\nu} + (D_\mu
\phi)(D^\mu\phi)^*
              -\lambda(|\phi|^2-c^2)^2  \label{eq:Lagrangian}
\end{equation}
 where the mass term for the photon has to be included in
order to give the
bosonized Lagrangian the correct symmetry. The properly
regularized vector
current is then
\begin{equation}
 J^\mu = \frac{1}{\sqrt{\pi}}
            \epsilon^{\mu\nu} \left(\partial_\nu \chi -
\frac{e}{\sqrt{\pi}}
A_\nu\right)
\end{equation}
which obeys the anomaly equation (\ref{eq:anomaly}).

The main reason for using the Bose-equivalent formulation is
that the anomaly
equation of the
 fermionic theory is present at the classical level, provided
the Lagrangian is
properly regularized.    This ensures that the level-crossing
effects connected
to the anomaly are included in the classical equations of
motion. This is
preferable to the usual analyses (e.g.\ see
\cite{Bochkarev,Carson,Turok})
where fermions are essentially ignored except for the fact
that, through the
anomaly equation, a change in the Chern-Simons number of a
given gauge field
configuration is known to be  accompanied by a corresponding
change in the
fermion number.

The static energy density is easily obtained in the $\chi =
A_0 = 0$ gauge
\cite{Roberge90}
\begin{equation}
 {\cal E} = \frac{e^2}{2\pi} A_1^2 + \left|\left(\frac{d}{dx}
-
ieA_1\right)\phi \right|^2
 +  \lambda(|\phi|^2 - c^2)^2.
\label{eq:energy}
\end{equation}
  As a result of this gauge
 choice, the fermion number and the Chern-Simons number are
identical.

Suppose one ignores the explicit mass term for the photon,
which is equivalent
to
 ignoring the contribution from the fermions.  One then finds
that $\cal E$ has
 an infinite number of local minima such that
\begin{eqnarray}
 A_1 &=& \frac{\pi N_{CS}}{eL} \nonumber\\
  \phi &=& c \exp(ieA_1 x)
\end{eqnarray}
where $N_{CS}$, an integer, is the Chern-Simons number.
Following \cite{Bochkarev,Carson}, one can construct a
noncontractible path
interpolating between  two vacuum states\cite{Manton} using
the
parametrization:
\begin{eqnarray}
 A_1(\tau) &=& \frac{\pi}{eL} ( N+\tau) \nonumber\\
 \phi(x; \tau) &=& c\exp(ieA_1 x)\left[ \cos\pi\tau + i
\sin\pi\tau\Phi(z)\right]
\label{eq:param}
\end{eqnarray}
where $z = \sqrt{\lambda}c^2\sin(\pi\tau )\,x$.
 As explained by Manton\cite{Manton} and Carson\cite{Carson},
the sphaleron
configuration is obtained through a minimax procedure as
follows:  the set of
path $\{\phi(x;\tau), A_1(\tau)\}$ are finite-energy field
configurations that
interpolate between two vacuum states as $\tau$ runs from 0 to
1.   Therefore,
as a function of $\tau$, there must exist a point along the
path where the
energy reaches a maximum, $E_{\rm max}$.  By considering the
set of all such
paths, one can find a function $\Phi(z)$ for which the energy
$E_{\rm max}$ is
minimal.  This configuration is the desired sphaleron which is
a solution to
the static equations of motion that corresponds to a saddle
point of the energy
functional for this system.
The importance of these sphaleron configurations is that they
are the main
contributor to
baryon-number violating processes occuring at finite
temperature\cite{Kuzmin}.

 It is straightforward to show that the sphaleron
configuration for the Abelian
Higgs model without fermions is given by $\Phi(z) = \tanh(z)$
and
$\tau=\frac{1}{2}$ in the
 limit as $L \rightarrow \infty$.  The energy along the
corresponding path is
then
\begin{equation}
 E(\tau) = \frac{8\sqrt{\lambda} c^3}{3}
\left|\sin^3\pi\tau\right|.
\end{equation}
This results in a periodic effective potential as function of
the fermionic
density (Chern-Simons number) as shown by the dashed line in
figure 1.

In order  to consider the effects of including the fermions in
the system, one
may reintroduce the mass term for the photon.
Naively, the corresponding effective potential is then as
illustrated by the
solid line in
figure 1\cite{Dine,Mottola,McLerran}, where the periodic
potential obtained
before is simply added to a pure fermionic contribution.
This results in potential barriers of varying height
separating the various
distinct local minima up to a maximum value for the fermion
density.
The computation of transition rates between a state having
$N$ fermions to a
state having
$N+1$ fermions is now more complicated since the transition is
between two
states having different energies rather than a true
vacuum-to-vacuum
transition.

Rather than computing the transition rate as a function of the
fermionic
density, suppose one simply wants to compute the critical
density.
Since the relevant states have a non-vanishing energy, it is
useful,
at finite density and in the $L \rightarrow \infty$ limit, to
subtract an
infinite constant from the energy and consider instead
\begin{equation}
 E(\tau) = \int^\infty_{-\infty} dx \left( {\cal E}(N, \tau) -
{\cal E}(N,
\tau=0)\right)
\end{equation}
Because the new effective potential was obtained by assuming
that one could
simply add two
contributions, it  may seem reasonable to assume that the same
function
$\Phi(z) = \tanh(z)$ found before still corresponds to the
sphaleron
configuration.
Thus, one finds
\begin{equation}
E(\tau) =  2\pi\tau n + \frac{8\sqrt{\lambda} c^3}{3}
\left|\sin^3\pi\tau\right|
\end{equation}
where $n = N/L$ is the finite fermion {\em density\/} taken to
be finite even
in the
$L \rightarrow \infty$ limit.
This result shows that it is possible to find a local extremum
of $E(\tau)$
only if $A_1$ does not exceed a critical value given by $32
\pi\lambda
c^3/3e\sqrt{3}$.

However, this result indicates that such a naive way of
deriving an effective
potential at finite density is not correct since, as has been
previously
found\cite{Rubakov2,Matveev87,Roberge90},
no local minima of the energy exists when $A_1$ exceeds
$2\sqrt{2} \pi\lambda c^3/3e\sqrt{3}$.
Furthermore, this derivation doesn't take into account the
fermion
back-reaction and  the function $\Phi(z)$ is {\em not\/} a
static solution to
the complete set of equations of motion of the original
Lagrangian (eq.
\ref{eq:Lagrangian}).  Writing $\phi = \Phi\exp(i\rho)$, the
static equations
of motion are:
\begin{eqnarray}
e A_1 - 2\pi\Phi^2 (\partial_x\rho - e A_1) &=& 0 \label{eom1}
\\
\partial_x \left[ \Phi^2(\partial_x\rho - e A_1)\right] &=& 0
\\
\partial_x^2\Phi - \Phi(\partial_x\rho - e A_1)^2 -
2\lambda\Phi (\Phi^2 - c^2)
&=& 0  \label{eom2}
\end{eqnarray}
Any solution to these equations has to be such that $A_1$ is
spatially constant
and that \begin{equation}
 \rho = e A_1 \int^x \left(1  + \frac{e}{2\pi\Phi^2} \right)
dx' \label{eq:rho}
\end{equation}
The periodicity requirement on $\rho$ then implies that the
various local
minima of the energy functional are such  that
\begin{equation}
N_{CS} = \frac{e}{\pi} A_1 L  =
 \frac{2NL}{\displaystyle \int^L_{-L}\left(\displaystyle 1  +
\frac{e}{2\pi\Phi^2}\right) dx}
\end{equation}
is no longer an integer.  In other words, the minima no longer
coincide with
pure gauge configurations.  Furthermore, the difference
between two adjacent
minima of this
 quantity is also different from unity.  This complicates the
search for a
non-contractible  path joining two adjacent minima since it
becomes very
difficult to find a parametrization consistent with the
periodic boundary
condition.  However, a static solution to the equations of
motion resembling
the Abelian Higgs sphaleron solution can still be found.
If one combines equations (\ref{eom1}) and (\ref{eom2}), one
gets
\begin{equation}
 \frac{d^2\Phi}{dx^2} = \frac{e^2A_1^2}{\pi^2\Phi^3} +
2\lambda\Phi(\Phi^2 -
c^2)
\end{equation}
which, in the $L\rightarrow\infty$ limit,  has a single
non-homogeneous
solution\cite{Roberge2}
as well as two homogeneous solutions.  To see this, it is
convenient to
parametrize $A_1$ in terms of a new variable $\gamma$ such
that
\begin{equation}
 A_1 = \left(1-\frac{\gamma}{3}\right)
\sqrt{\frac{2\gamma\lambda}{3}}
\frac{\pi c^3}{e}\qquad,
\qquad 0 \le \gamma \le 1
\end{equation}
This parametrization is one-to-one on the defined interval and
such that
$A_1(\gamma=0)=0$ while $A_1(\gamma=1)$ is the critical
density above which no
local minima of the energy exists.
With this parametrization, the inhomogeneous solution
(sphaleron) is given by
\begin{equation}
 \Phi_{sph} = c \sqrt{\frac{2\gamma}{3} + (1-\gamma)
\tanh^2\left(\sqrt{\lambda(1-\gamma)}cx\right)}
\end{equation}
and the two homogeneous solutions are
\begin{eqnarray}
 \Phi_1 &=& c\sqrt{1 - \frac{\gamma}{3}} \nonumber\\
 \Phi_2 &=& \frac{c}{\sqrt{6}} \sqrt{\gamma +
\sqrt{3\gamma(4-\gamma)}}
\end{eqnarray}
The difference in energy between the sphaleron solution and
the lowest energy
 homogeneous solution ($\Phi_1$) is
$8c^3\sqrt{\lambda(1-\gamma)}/3$ which
correctly reproduces the Abelian Higgs result when the fermion
density is zero
($\gamma=0$), vanishes at the critical density ($\gamma=1$),
and remains finite
for all other values of $\gamma$.  The result for the
effective potential is
given in Figure 2.
There are three important characteristics of this effective
potential; firstly,
its overall positive curvature due to the finite fermionic
density; secondly
the existence of a critical density above which no stable
solution can be
found; and lastly, and perhaps most importantly, the decrease
of the sphaleron
energy (the barrier height) with increasing fermion number.
 Of course, the important question is the extrapolation of
these results for
3+1 dimensional theories.
 Firstly, in order to compute transition rates at finite
temperature, the
relevant
 quantity is the free energy which also has a generally
positive curvature as a
function of the fermion
density\cite{Khlebnikov,Dine,Mottola,McLerran,footnote}.
 Secondly, in 3+1 dimensional models there also exists a
critical
density\cite{Rubakov2,Matveev88}, just as in the $1+1$
dimensional model
discussed above.
 Hence, it is probably safe to assume that the height of the
barrier between
local minima
 (the energy of the sphaleron) decreases as the fermionic
density increases
 so that it vanishes at and above the critical density.
 From this it follows that, as the fermionic density
increases, sphaleron
induced transitions would occur even more rapidly than the
rate suggested by
conventional calculations, with the finite density effects
becoming the
dominant factor when the fermionic density becomes comparable
to or greater
than the critical density.
The critical density computed in the standard
model\cite{Rubakov2,Matveev88} is
about 12 orders of magnitude greater than the nuclear density
which is
comparable to the total energy density close to the
electroweak phase
transition in the standard Big Bang model\cite{Kolb}. However,
the critical
density is the {\em net} fermionic density (i.e.\  fermion
minus anti-fermion)
which is likely to be much smaller than the overall density
given the small
baryon to photon ratio obtained from primordial
nucleosynthesis.  Thus it
appears that the rate increase of sphaleron transitions due to
finite density
effects would be negligible.  Nonetheless, it might still be
desirable to
compute this rate increase precisely and to do so it will be
necessary to know
in what way the standard  sphaleron\cite{Manton2} is modified
at finite
density.   It is likely that non-perturbative effects would
play a crucial role
thereby leading to an impossibly complicated analytical
solution.  However, a
numerical solution should be attainable.

\acknowledgments

J'aimerais remercier le Conseil de l'Enseignement
Franco-Ontarien pour sa
contribution financi\`ere \`a ce projet, ainsi que R. Haq pour
une discussion
tr\`es utile.  I would also like to thank the referees for
their useful
suggestions.

\begin{figure}
\caption{Schematic representation of the naive effective
potential as a
function of the fermion density.  The dashed line represents
the abelian Higgs
contribution which has an infinite number of local minima.
The dotted line is the pure fermion contribution.
The solid line, obtained  by adding the abelian Higgs
sphaleron contribution to
the pure fermion contribution is {\em not} the true effective
potential.}
\vskip 1cm

\caption{Schematic representation of the real effective
potential as a function
of the fermion density.
The main thing to note is that the height of the potential
barrier decreases
with increasing fermion density until it vanishes at the
critical density.}
\end{figure}


\begin{references}
\bibitem{Roberge2} Strictly speaking there are many other such
solutions but,
apart from possible translations, the solution given is the
one with the lowest
energy of all inhomogeneous solutions.  The method used to
derive this solution
as well as a thorough analysis of
this model on a circle of finite diameter are given in A.
Roberge (in
preparation).
\bibitem{footnote}
Note  that in the 1+1 dimensional model we introduced, the
static energy is
really just the same as the zero temperature free energy.
\end{references}
\end{document}